\documentclass[12pt]{article}

\usepackage{amsmath}
\usepackage{amsfonts}
\usepackage{amssymb}
\usepackage{amsthm}
\usepackage{setspace}
\usepackage{color}
\usepackage{subcaption}
\usepackage{caption}
\usepackage{enumerate}
\usepackage[hidelinks]{hyperref}
\usepackage{graphicx}
\usepackage[hidelinks]{hyperref} 
\usepackage[nottoc]{tocbibind}
\usepackage[utf8]{inputenc}

\usepackage{titlesec}

\titleformat*{\section}{\large\bfseries}

\newcommand\ee{\end{equation}}
\newcommand\be{\begin{equation}}
\newcommand\eea{\end{eqnarray}}
\newcommand\bea{\begin{eqnarray}}

\newcommand\mpl{M_{\rm pl}}
\newcommand\comment[1]{}

\comment{
\hypersetup{
colorlinks=true,
citecolor=DarkBlue,
linkcolor=DarkBlue,
urlcolor=DarkBlue,
}}

\def\O{\mathcal{O}}
\def\tr{{\rm Tr}}
\def\d{\partial}

\def\L{{\mathcal L}}
\def\LEH{\L_{\rm EH}}
\def\vep{\varepsilon}

\def\vphi{\varphi}

\def\ep{\epsilon}

\def\nmax{n_{\rm max}}
\def\A{{\mathcal A}}

\begin{document}

\begin{center}

  {\Large\bf How Heavy Can Moduli Be?}

\vskip 1 cm
{Mehrdad Mirbabayi$^a$ and Giovanni Villadoro $^{a,b}$}
\vskip 0.5 cm

{\em $^a$ Abdus Salam International Centre for Theoretical Physics, \\ Strada Costiera 11, 34151, Trieste, Italy}

{\em $^b$ INFN, Sezione di Trieste, Via Valerio 2, I-34127 Trieste, Italy}

\vskip 1cm

\end{center}
\noindent {\bf Abstract:} {\small In Kaluza-Klein (KK) compactification of gravitational theories, moduli fields, which are scalar fields associated to the deformations of the compact manifold, are {\em typically} lighter than the KK gravitons. However, a universal limit on their mass does not seem to exist. We provide numerical evidence that a light scalar particle, with mass ratio to the first KK graviton $(m_{\rm sc}/m_{1KK})^2 \leq {4/3}$, is necessary for the consistency of the $4d$ effective theory of KK gravitons. This can be interpreted as a limit on how rigidly the compact manifold can be stabilized.}

\vskip 1 cm

\section{Introduction}

Presence of light scalar moduli is a generic feature of Kaluza-Klein theories. The size and shape of the internal manifold are gravitational degrees of freedom, which need to be stabilized via a balance among different sources of stress and energy. Two such examples are the Goldberger-Wise model, with a varying bulk scalar field \cite{GW}, and the loop-level stabilization with several Casimir contributions \cite{Arkani}. At low back-reaction, the lightest modulus is typically much lighter than the KK scale.

It is interesting to ask how rigid the compact manifold can be made, perhaps by going to the strong back-reaction regime or considering hyperbolic internal manifolds \cite{Kaloper}. More specifically, can the mass of the lightest KK graviton, $m_{1KK}$, be parametrically lower than the lightest scalar field? Our goal in this paper is to study this question, not via a survey of compactification scenarios and stabilization mechanisms, but using the effective theory of KK gravitons.

In fact, there are arguments that suggest the theory of an isolated massive spin-2 particle, namely a particle whose mass is parametrically smaller than the next heavy particle in the spectrum, does not admit a conventional UV completion \cite{Bellazzini}. However, the tower of KK gravitons automatically fills the gap, whether or not there is a light scalar. To say something about the lightest scalar field, we use the extra input that the KK theory is supposed to remain weakly coupled at energies well above $m_{1KK}$ and it has to reproduce the predictions of the Einstein theory in the $D$-dimensional bulk. Of course, the question is sensible only if $m_{1KK}\ll M_D$, the bulk Planck mass. In the intermediate range $m_{1KK}\ll E\ll M_D$, the bulk scattering amplitudes are dominated by the tree-level diagrams and they grow $\propto E^2$. The same must hold in the KK theory. 

It is well-known that the tree-level scattering amplitudes in a theory of a single massive spin-2 particle have a much steeper UV behavior. Generically, the amplitude for longitudinal polarizations grows as $E^{10}$
\cite{Arkani}, while a judicious choice of self-couplings {\em a la} dRGT \cite{dRGT1,dRGT2} softens it to $E^6$. The coupling between different particles in the KK theory is therefore essential to reproduce the bulk expectation. Indeed, we find that the presence of a light scalar particle is necessary.

A related question can be (and it has been) asked about the theory of massive non-Abelian gauge bosons. Here again the scattering amplitudes containing the longitudinal polarizations have a hard UV behavior, becoming strong at $m/g$. The standard way of UV completing the model -- the Higgs mechanism -- involves a scalar field. However, one could ask if it is possible to push the strong coupling scale parametrically above $m/g$ without introducing a scalar field. In fact, compactification of Yang-Mills theory in one higher dimension can result in such a ``Higgs-less Electroweak'' model \cite{Chivukula,Csaki}. Here, it is the interaction among the various KK vectors, without the interference of any scalar field, that softens the UV behavior of the amplitude. 

There is also a purely field theoretic (non-gravitational) way of formulating our question. Large-$N$ confining theories generically lead to a spectrum of massive mesons, which become strongly coupled at an energy parametrically larger than their mass. This parametric separation and dispersion relations have been used to learn about the spectrum and the low-energy couplings in such theories (see e.g. \cite{Albert,Fernandez} for interesting recent progress). One could ask if it is possible in such a scenario to make the lightest spin-2 meson parametrically lighter than the lightest spin-0 one. Our analysis suggests a negative answer, albeit with different assumptions, namely the absence of spin $>2$ particles and a UV growth no faster than $E^2$.

The two interpretations, field theoretic and gravitational, are related by the fact that in models of warped compactification the massless graviton can be made arbitrarily weakly coupled to the rest of the spectrum, leaving us with a non-gravitational $4d$ QFT that becomes strongly coupled at a parametrically larger energy scale than its gap $M_D\gg m_{1KK}$ \cite{AdS/CFT}. Our analysis has been limited to Einstein gravity in the bulk, or large 't Hooft coupling at in boundary. It would be interesting to generalize it to include higher spin particles in the spectrum, or equivalently the stringy effects in the bulk, as a more realistic model of large-$N$ confining theories. 

\subsection{General Estimates}

In this work, we consider only the elastic 2-to-2 scattering of the lightest massive spin-2 particle ($1KK$) at tree level. This depends on the spectrum of bosonic particles (masses and spins), their cubic coupling to $1KK$, and the quartic self-coupling of $1KK$. We assume the bulk theory is Einstein gravity, Supergravity, or any weakly coupled field theory coupled to gravity. So the maximum spin is 2. 

The amplitude is of the form
\be
\A_{\rm tree} = \A_{\rm exchange} + \A_{\rm contact}.
\ee
At large center of mass energy ${E}$, the maximum growth rate of each contribution depends on the external polarizations. The helicity-0 ($S$) and helicity-1 ($V$) polarization tensors grow respectively as ${E}^2$ and ${E}^1$, while helicity-2 ($T$) polarizations do not depend on ${E}$. Likewise the propagators of massive spin-2 particles grow as ${E}^2$. The explicit forms of these kinematic data are reviewed in section \ref{sec:kin}.

With interactions of at most two derivatives, the asymptotic behavior of the contact term at a generic scattering angle is expected to be $\A_{\rm contact} \propto {E}^{2n_{\rm max}}$, where $n_{\rm max} = 5$ for the fully longitudinal scattering ($SSSS$), 4 for $SVSV$, 3 for $VVVV$ and so on. The exchange contribution is naively expected to grow faster, as ${E}^{2 n_{\rm max}+4}$ because of one extra vertex and a propagator. However, if the structure of 2-derivative interactions are fixed by the Einstein theory (as is the case in Kaluza-Klein theory) then we also have $\A_{\rm exchange} \propto {E}^{2n_{\rm max}}$. 

Our goal is to find the conditions on the spectrum and couplings under which the UV behavior of the tree-level amplitude softens to $A_{\rm tree}\propto {E}^2$. Expanding in powers of ${E}$, the contact amplitude has the following schematic structure
\be
A_{\rm contact}^{(2n)} \sim \frac{\mu_{1111} {E}^{2n}}{f^2 m_1^{2(n-1)}},\qquad n\leq \nmax
\ee
where $m_1 = m_{1KK}$, and $f$ is an overall energy scale to make couplings such as $\mu_{1111}$ dimensionless. For non-warped compactifications, it is related to the $4d$ gravitational interaction strength: $f^2 = \mpl^2$. The exchange amplitude has the form
\be
A_{\rm exchange}^{(2n)} \sim \sum_{i} \mu_{11i}^2 \frac{{E}^{2n}}{f^2 m_1^{2(n -s_i- 1)}m_i^{2 s_i}}P_{i,n}((m_i/m_1)^2),
\ee
where the sum is over the intermediate particles with mass $m_i$ and spin $s_i$, the factor $1/m_i^{2 s_i}$ comes from the leading high energy behavior of the propagator, and $P_{i,n}$ is a polynomial of maximum degree $\nmax +2s_i-n-2$ coming from the subleading terms. In the case of the graviton exchange, $m_i=0$ and $P_{i,n}$ degenerates into a monomial that cancels powers of $m_i$.

Of course, generically there is more than one interaction vertex among any subset of particles, and each amplitude has a nontrivial dependence on the scattering angle. We review the interaction vertices in section \ref{sec:int}. In section \ref{sec:cons}, we derive 13 independent constraints by requiring the cancellation of terms that grow faster than $E^2$. Many of these constraints involve infinite sums over cubic couplings, weighted by various powers of $m_i/m_1$. We discuss the numerical solutions obtained by truncating these sums, and the resulting mass threshold, in section \ref{sec:sol}, and conclude in section \ref{sec:con}.

\section{Kinematics}\label{sec:kin}

Our scattering setup is similar to the one studied in \cite{Cheung}. The external particles are four $1KK$ modes with mass $m_1$. Let us define a family of on-shell momenta
\be
k^\mu (\alpha) = ({E},\kappa \sin\alpha,0,\kappa \cos \alpha),
\ee
where $\kappa = \sqrt{{E}^2-m_1^2}$. We choose the external ``in-going'' momenta to be
\be
k_1^\mu =k^\mu(0),\quad k_2^\mu =k^\mu(\pi), \quad k_3^\mu=-k^\mu(\alpha), \quad k_4^\mu=- k^\mu(\pi-\alpha).
\ee
External polarizations satisfy $k^\mu \vep_{\mu\nu} =0= \eta^{\mu\nu}\vep_{\mu\nu} $. They can be constructed using the spatially transverse and longitudinal polarization vectors
\be\begin{split}
e^L_\mu(\alpha) &= \frac{1}{m_1}(\kappa, {E} \sin \alpha,0,{E} \cos \alpha),\\[10pt]
e^\pm_\mu(\alpha) &= \frac{1}{\sqrt{2}} (0,\cos \alpha,\pm i,-\sin \alpha),\end{split}
\ee
There are two helicity-2 (tensor) polarizations 
\be
\vep^{\pm \pm} _{\mu\nu}(\alpha) = e^\pm_\mu(\alpha) e^\pm_\nu(\alpha),
\ee
two helicity-1 polarizations
\be
\vep^{L\pm} _{\mu\nu}(\alpha) = \frac{1}{\sqrt{2}} [e^L_\mu(\alpha) e^\pm_\nu(\alpha)+e^L_\nu(\alpha) e^\pm_\mu(\alpha)],
\ee
and one helicity-0, fully longitudinal, polarization
\be
\vep^{L} _{\mu\nu}(\alpha) = \sqrt{\frac{3}{2}}
\left[e^L_\mu(\alpha) e^L_\nu(\alpha)-\frac{1}{3} \Pi_{\mu\nu}(k(\alpha),m_1)\right],
\ee
where in our mostly plus metric convention the transverse projector is defined as
\be
\Pi_{\mu\nu}(p,m) = \eta_{\mu\nu}+\frac{p_\mu p_\nu}{m^2}.
\ee
We will also need the propagators for the exchanged particles. The massive spin-2 propagator is
\be
P^{\rm KK}_{\mu\nu\alpha\beta}(p,m) = \frac{i}{2(p^2+m^2)}\left[\Pi_{\mu\alpha}(p,m)\Pi_{\nu\beta}(p,m)+\Pi_{\mu\beta}(p,m) \Pi_{\nu\alpha}(p,m)-\frac{2}{3}\Pi_{\mu\nu}(p,m)\Pi_{\alpha\beta}(p,m)\right],
\ee
the graviton propagator 
\be
P^{\rm gr}_{\mu\nu\alpha\beta}(p) = \frac{i}{2p^2}\left[\Pi^0_{\mu\alpha}(p)\Pi^0_{\nu\beta}(p)+\Pi^0_{\mu\beta}(p) \Pi^0_{\nu\alpha}(p)-\Pi^0_{\mu\nu}(p)\Pi^0_{\alpha\beta}(p)\right],
\ee
where
\be
\Pi^0_{\mu\beta}(p) = \eta_{\mu\nu}+\xi\frac{p_\mu p_\nu}{p^2},
\ee
and $\xi$ is an arbitrary gauge parameter, the propagator for massive vectors
\be
P^{\rm V}_{\mu\nu}(p,m) = \frac{i\Pi_{\mu\nu}(p,m)}{p^2+m^2},
\ee
and that of scalars is simply $i/{(p^2+m^2)}$.

\section{Interactions}\label{sec:int}

Compactification of Einstein theory results in coupling between spin-2 fields with at most two derivatives, coupling between spin-1 and spin-2 fields with at most one derivative, and the coupling between scalar and spin-2 fields with zero derivatives. Moreover, the 2-derivative interactions have the same structure as in Einstein theory. More specifically, consider the $4d$ Einstein-Hilbert Lagrangian:
\be
\LEH =\frac{f^2}{2}\sqrt{-g} R.
\ee
The structure of 2-derivative interactions between KK gravitons $h^{(i)}_{\mu\nu}$ (with $i=0$ corresponding to the massless graviton, $i = 1$ to the $1KK$, etc.) are obtained by substituting $g_{\mu\nu} = \eta_{\mu\nu}+\sum_{i}\ep_i h_{\mu\nu}^{(i)}$ and Taylor expanding in $\ep_i$ and picking the appropriate power. In particular, the cubic self-coupling of $1KK$ is
\be
\L^{{\rm 2-der}}_{111} = \mu_{111} \frac{1}{3!}\frac{\d^3 \LEH}{\d \ep_1^3}\Big|_{\{\ep_{j}\} = 0},
\ee
and the cubic coupling between two $1KK$ and another mode is
\be
\L^{{\rm 2-der}}_{11i} = \mu_{11i} \frac{1}{2}\frac{\d^3 \LEH}{\d \ep_1^2\d \ep_i}\Big|_{\{\ep_{j}\} = 0},
\ee
where $\mu_{111}$ and $\mu_{11i}$ are dimensionless coefficients that depend on the details of compactification. Similarly, the quartic self-coupling is
\be
\L^{{\rm 2-der}}_{1111} = \mu_{1111} \frac{1}{4!}\frac{\d^4 \LEH}{\d \ep_1^4}\Big|_{\{\ep_{j}\} = 0}.
\ee
The above fixed structures are inherited from the higher dimensional Einstein-Hilbert action because all derivatives are taken to be along the non-compact $4d$ dimensions. The coefficients for different combinations of the KK modes differ because each KK mode has a different profile $f^{(i)}(\vec y)$ in the compact dimensions $\vec y$. But all 2-derivative interactions that contain the same combination of the KK modes multiply the same integral over the compact manifold. 

At zero derivative level, the most general cubic interaction of a tensor $h_{\mu\nu}$, written in matrix notation, is
\be
\L^{\rm 0-der}_3 = -\frac{f^2 m^2}{8} \left(c_1 \tr[\eta\cdot h\cdot \eta\cdot h\cdot \eta\cdot h]
+c_2  \tr[\eta\cdot h\cdot \eta\cdot h] \tr[\eta\cdot h]
+c_3 \tr[\eta\cdot h]^3\right).
\ee
The cubic self-coupling of $1KK$ can be obtained by substituting $h_{\mu\nu}\to h^{(1)}_{\mu\nu}$, $m\to m_1$ and $c_{a} \to \mu_{111} c_{a}^{111}$ for $a=1,2,3$. Here, the explicit factor of $\mu_{111}$ is to simplify future relations. The cubic coupling of two $1KK$ modes with the mode $i\neq 1$ is obtained by $h_{\mu\nu}\to \ep_1 h^{(1)}_{\mu\nu}+\ep_i h^{(i)}_{\mu\nu}$, expanding to order $\ep_1^2 \ep_i$ and replacing $m\to m_1$ and $c_a \to \mu_{11i} c_a^{11i}$. Again $\{c_a^{ijk}\}$ depend on the details of the compactification.

We note that $c_3$ will not contribute to the tree-level scattering of $1KK$, because the external polarizations are traceless. Moreover, the cubic coupling to the massless graviton is fixed by diffeomorphysm invariance (no dependence on $\xi$)
\be
c_1^{110} = -\frac{2}{3},\qquad c_2^{110} = \frac{1}{2}.
\ee
In writing the quartic self-coupling of $h^{(1)}_{\mu\nu}$, we only demonstrate the structures that do contribute to the tree-level on-shell amplitude
\be
\L^{\rm 0-der}_4 = -\mu_{1111}\frac{f^2 m_1^2}{8} \left(d_1 \tr[\eta\cdot h^{(1)}\cdot \eta\cdot h^{(1)}\cdot \eta\cdot h^{(1)}\cdot \eta\cdot h^{(1)}]
+ d_3 \tr[\eta\cdot h^{(1)}\cdot \eta\cdot h^{(1)}]^2+ \cdots\right). 
\ee
To be general, we include the interaction between a massive vector $V_\mu$ and two $1KK$ particles though we do not know any example where such a coupling arises. Restricting to at most 2 derivatives, the only possible such interaction will have one derivative and can be obtained from the 0-derivative cubic coupling between $h^{(1)}_{\mu\nu}$ and $h^{(i)}_{\mu\nu}$ by replacing $h^{(i)}_{\mu\nu}\to \d_\mu V_\nu + \d_\nu V_\mu $. We will not include a massless vector field because if $1KK$ is electrically charged, we can choose a basis in which the photon exchange will not contribute to the diagonal matrix elements. 

Finally, the cubic 0-derivative interaction between a scalar particle and two $1KK$ particles can be obtained by replacing $h^{(i)}_{\mu\nu}\to \eta_{\mu\nu} \vphi$. Clearly, only the linear combination $c_\vphi = c_1^{11\vphi} + \frac{4}{3} c_2^{11\vphi}$ will appear in the 1KK 2to2 scattering. We could also write 2-derivative vertices, but they do not seem to arise from compactification, and in any case they do not help to satisfy the constraints. 

\section{Constraints}\label{sec:cons}

We find that requiring no faster than $E^2$ growth in the following amplitudes 
\be
SS\to SS,\quad  SV\to SV,\quad V^+ V^- \to V^+ V^-,\quad  V^+ V^+ \to V^+ V^+,
\ee
guarantees that also the amplitudes involving tensor modes are soft. These amplitude are even in ${E}$ and $\alpha$ and at every order up to ${E^4}$, the $\alpha$ dependence is a cosine series that terminates at $\cos(4\alpha)$ or earlier. (The large $E$ expansion can lead to singularities in the forward limit, but because the maximum exchanged spin is 2, they appear at $\O(E^2)$ or below.) There is a total of 13 independent constraints. To express them in a concise way we make the followiing definitions.

Take $g_i = {\mu_{11i}}/{\sqrt{\mu_{1111}}}$ and the array of products
\be
T_i = g_i^2\times \{ c_{1}^{11i}, c_{2}^{11i}, (c_1^{11i})^2, (c_2^{11i})^2, c_1^{11i} c_2^{11i},1 \},
\ee
then the following sums over the spin-2 data appear in the constraints
\be\label{X}
X[a,b] = \sum_{i=1}^\infty  T_i[a] \ m_i^{2 b- 6},\qquad 1\leq a,b\leq 6,
\ee
where all masses here and below are measured in units of $m_1$. Similarly, normalizing the cubic couplings between $1KK$ and massive spin-1 particles as
\be
V_i = \frac{1}{\mu_{1111}}\times \{{(c_{1}^{11V_i})^2},{(c_{2}^{11V_i})^2},{c_{1}^{11V_i} c_{2}^{11V_i}}\},
\ee
the following sums over the spin-1 data appear
\be\label{W}
W[a,b] = \sum_{i=1}^\infty  V_i[a] \ m_{V_i}^{2 b- 4},\qquad 1\leq a,b\leq 3.
\ee
We also normalize the coupling of $1KK$ to the graviton and scalar as
\be
c_{\rm gr} = \frac{\mu_{110}}{\sqrt{\mu_{1111}}},\qquad c_{\rm sc} =\frac{c_\vphi}{\sqrt{\mu_{1111}}}.
\ee
The constraints are of the form
\be\label{constraints}
M_X \cdot \vec X + M_W \cdot \vec W + \vec b =0,
\ee
where $\vec X[n]= X[\lceil \frac{n}{6}\rceil,(n-1)\pmod{6}+1]$, and $\vec W[n]= W[\lceil \frac{n}{3}\rceil,(n-1)\pmod{3}+1]$, and $M_X$ and $M_W$ are sparse matrices whose only nonzero elements are
\be\begin{split}
\Big\{& M_X^{{1,1}}= 1,M_X^{{1,19}}= -\frac{4}{3},M_X^{{1,20}}= -\frac{8}{3}, M_X^{{1,25}}=
   -2,M_X^{{1,26}}= -4,M_X^{{1,31}}= \frac{1}{3},M_X^{{1,32}}= -\frac{2}{3},\\ & M_X^{{2,2}}=
   1,M_X^{{2,32}}= \frac{2}{3},M_X^{{3,3}}= 1,M_X^{{4,4}}= 1,M_X^{{4,35}}=
   \frac{5}{12}, M_X^{{5,5}}= 1,M_X^{{5,16}}= 3,M_X^{{5,35}}= -\frac{7}{4},\\&  M_X^{{5,36}}=
   \frac{1}{12}, M_X^{{6,7}}= 1,M_X^{{6,19}}= -1,M_X^{{6,20}}= -3,M_X^{{6,26}}=
   -\frac{9}{2}, M_X^{{6,31}}= -\frac{1}{4},M_X^{{6,32}}= -\frac{3}{4},\\& M_X^{{7,8}}=
   1,M_X^{{7,20}}= 2, M_X^{{7,26}}= 3,M_X^{{8,13}}= 1,M_X^{{8,19}}= \frac{16}{9},M_X^{{8,20}}=
   \frac{32}{9}, M_X^{{8,25}}= \frac{8}{3},M_X^{{8,26}}= \frac{16}{3},\\& M_X^{{8,32}}=
   \frac{8}{9}, M_X^{{9,14}}= 1,M_X^{{9,32}}= -\frac{4}{9},M_X^{{10,15}}= 1,M_X^{{10,35}}=
   -\frac{1}{9}, M_X^{{11,33}}= 1,M_X^{{12,34}}= 1\Big\}
   \end{split}
\ee
\be\begin{split}
   \Big\{& M_W^{{1,2}}= 81,M_W^{{1,4}}= -8,M_W^{{1,7}}= -12,M_W^{{2,2}}= -12, M_W^{{3,2}}=
   -8,M_W^{{4,2}}= -24,M_W^{{4,3}}= -\frac{18}{5},\\ & M_W^{{5,2}}= \frac{941}{5},
   M_W^{{5,3}}=
   \frac{126}{5},M_W^{{6,2}}= \frac{567}{8},M_W^{{6,4}}= -6,M_W^{{7,2}}=
   -\frac{189}{4}, M_W^{{8,1}}= 6,M_W^{{8,2}}= -100,\\ &  M_W^{{8,4}}= \frac{32}{3},M_W^{{8,7}}=
   16,M_W^{{9,2}}= \frac{86}{3},M_W^{{10,2}}= \frac{424}{15},M_W^{{10,3}}=
   \frac{6}{5}, M_W^{{12,2}}= -12,M_W^{{13,2}}= 18\Big\}
   \end{split}
\ee
and the vector $\vec b$ is 
\be\begin{split}
   \vec b = \Big\{&\frac{1}{12} (5 c_{\rm gr}^2-96 d_3-37),\frac{1-c_{\rm gr}^2}{3},-\frac{2}{9}
   (3 c_{\rm gr}^2-5),\frac{1}{15} \left(-54 
   c_{\rm sc}^2+c_{\rm gr}^2+12\right),\\& \frac{1}{90} \left(81   c_{\rm sc}^2
   \left(5 m_{\rm sc}^2+32\right)-48 c_{\rm gr}^2-56\right),\frac{3}{16} (c_{\rm gr}^2-48
   d_3-17),\frac{1}{4} (c_{\rm gr}^2+24 d_3+7),\\& -\frac{4}{9} (c_{\rm gr}^2-24
   d_3-9),\frac{4}{27} (3 c_{\rm gr}^2-4),\frac{4}{27} (3
   c_{\rm gr}^2-7), c_{\rm gr}^2-1,-\frac{4}{3},d_1-1\Big\}
   \end{split}
\ee
To generalize the system \eqref{constraints} to include multiple scalars, for each scalar one can simply add a new coupling analogous to $c_{\rm sc}$ and its mass analogous to $m_{\rm sc}$. 
\section{Numerical Solution}\label{sec:sol}
Ideally, one would like to prove that \eqref{constraints} has no solution if $c_{\rm sc} = 0$. We could not show this analytically. Insead, we searched for solutions numerically. Namely, first we truncated the infinite sums so that we have a finite number of variables. Then we summed the squares of the first 12 constraints (the 13th one is easily solved to eliminate $d_1$), and looked for the minimum using the python function \texttt{scipy.optimize.basinhopping} with the local minimizer \texttt{"L-BFGS-B"}. For the numerical search, was used $\O(1)$ bounds on the derivative couplings, and $O(m_i^2/m_1^2)$ for non-derivative couplings $c^{11i}_a$, as expected from explicit compactifications. Our threshold for declaring the existence of a numerical solution is reaching a minimum below $10^{-10}$, with variables away from the upper or lower limits. Of course, we cannot rule out the possibility that a solution exists but it is numerically missed because of a smaller basin of attraction.

In practice, little was gained by adding more than two KK gravitons, or adding massive vector bosons or the massless graviton. Numerical search becomes less efficient by enlarging the space of variables. With no scalar field, we never found a minimum less than $10^{-3}$. Similarly, with 1 scalar field and only the first KK mode $1KK$, we did not find a minimum below $10^{-2}$. However, with 1 scalar and the first two KK modes, there exists a numerical solution as longs as $m_{\rm sc} <m_{\rm max}\approx 1.15 m_{1}$. As $m_{\rm sc}$ is increased from $0$ to $m_{\rm max}$, the second KK mass $m_2$ decreases from $\approx 2 m_1$ until it becomes degenerate with $m_{\rm sc}$ at $m_{\rm max}$. At this point the cubic self-coupling of $1KK$ vanishes, $g_1\to 0$, and we can find an analytic solution
\be
m_{\rm max}^2 = \frac{4}{3}m_1^2,\quad g_2 = 1,\quad c_{\rm sc}^2 = \frac{4}{243},\quad c_1^{112}=- \frac{10}{9},\quad c_2^{112}= \frac{3}{2},
\quad d_3 =  -\frac{5}{12}.
\ee
Adding more KK modes, more scalars, massive vector bosons, or the massless graviton made little difference in the above conclusion. 

\section{Conclusion}\label{sec:con}
In this work, we considered KK theories with a large hierarchy between the compactification scale and the bulk Planck scale $m_{1KK}\ll M_D$. As such, they have a weakly coupled $4d$ effective description with a tower of massive spin-2 KK gravitons and couplings that depend on the specifics of the compactification.

By demanding that the scattering amplitude of the first KK graviton should grow no faster than $E^2$ at high energies, as expected from Einstein gravity in the bulk, we inferred that the interaction among the tower of KK gravitons is not enough and there must exist a scalar particle in the spectrum no heavier than $\sqrt{4/3}\ m_{1KK}$. In explicit examples, the light moduli fields are indeed responsible for softening the KK graviton amplitudes from $E^6$ to $E^2$. Hence, our result suggests a limit on the stiffness of moduli stabilization. 


\vspace{0.3cm}
\noindent
\section*{Acknowledgments}

We thank Lorenzo Di Pietro, Sergei Dubovsky, John March-Russell, Riccardo Rattazzi for useful discussions. We also thank Nima Arkani-Hamed for posing the question to one of us 20 years ago.

\bibliography{bibrad}
\end{document}